# DESIGN AND DEVELOPMENT OF NOVEL ELECTROPLATING SPRING FRAME MEMS STRUCTURE SPECIMENS FOR THE MICROTENSILE TESTING OF THIN FILM MATERIALS

*Ming-Tzer Lin, Chi-Jia Tong & Chung-Hsun Chiang*

Institute of Precision Engineering, National Chung Hsing University, Taichung, 402, Taiwan
Email: mingtlin@nchu.edu.tw

## ABSTRACT

A novel design electroplating spring frame MEMS structure specimen is demonstrated here. The specimen can be fit into a specially designed microtensile apparatus which is capable of carrying out a series of tests on sub-micro scale freestanding thin films.

Certain thin films applicable in MEMS were tested including sputtered Chrome, Nicole, Copper, Tantalum Nitride thin films. Metal specimens were fabricated by sputtering; for Tantalum Nitride film samples, Nitrogen gas was introduced into the chamber during sputtering Tantalum films on the silicon wafer.

We found the modulus of Copper, Tantalum Nitride and 5% Au-Cr thin films thin films with thickness of 200 to 800nm at room temperature. The results demonstrated the yield stresses of metal films follow the trend of the Hall-Petch relation.

## 1. INTRODUCTION

Microelectromechanical systems (MEMS) technologies are developing rapidly with increasing study of the design, fabrication and commercialization of microscale systems and devices. Accurate mechanical properties of thin films are important for successful design and development of MEMS.

One possibility for measuring the stress-strain characteristics of materials for MEMS is to impose strain on a sample and determine the resultant stress. Mechanical properties testing of bulk materials are usually done by external actuation, in which a sample is pulled while its response is recorded.

For thin film materials application for MEMS, an alternative actuation option exists. Through the use of micromachining, internal actuation of thin films can be achieved. Some examples include bridge pull-down experiments [1] as well as resonance measurements using electrostatic comb drives [2]. Advantages of internal actuation include ease of testing once the structures have been fabricated, the possibility of fabricating a large number of test structures in parallel, and the resultant large number of tests that can be performed.

However, the primary drawbacks of the internal actuation are that calibration of the tests is not straightforward, and that accurate stress and strain values cannot easily be extracted. Also, since the available forces from electrostatic actuation are rather small, these tests are usually run at resonance to achieve larger displacements [3]. Thus a different test amplitude or test frequency will require a design change. Due to the complexity of the micromachining process involved in fabricating internally actuated test samples, the choice of materials that can be tested is fairly limited.

For external actuation, where the thin film sample is gripped and acted upon by applied deformation, less complicated sample designs can be used, and fewer restrictions on materials choices are imposed. Also, stresses and strains are measured externally and calibration is thus more straightforward. In addition, the cost of the sample design and fabrication can be reduced dramatically.

Despite that, it is not commonly applied due to the difficulties in sample handling and having the force transfer to the sample. Besides, strain determination at very small dimensions is another issue which needs to be considered.

Here, a novel setting that has the merit to eliminate such shortcoming and discrepancy has been demonstrated. It is designed to use electroplating spring frame MEMS structure specimen integrates pin-pin align holes, misalignment compensate spring structure frame, load sensor beam and freestanding thin film. The specimen is then fit into a specially designed microtensile apparatus to carry out a series of experimental micromechanical uniaxial stress tests.

## 2. EXPERIMENTAL

The experiments used a displacement-based monotonic loading of MEMS thin film samples with uniaxial microtensile testing approach. The materials choose here were typically applicable as structure or motion gears in MEMS.

The detail can be described into two categories:

### 2.1. SAMPLE DESIGN AND FABRICATION

The test chip is designed follows that of [4], [5] with major dimension and fabrication processes change (Figure 1,2,3) to fit in an apparatus show in Figure 6 where one end of the chip is pulled by a High Voltage Piezoelectric translator. The measured force was converted to nominal stress by dividing by the initial cross-sectional area of the film. The movement profile can be designed pending on all kinds of mechanical tests.

This experimental set-up consists of the testing chip fabricated from standard clean room processing containing the freestanding specimen, a force sensing mechanism, and an actuation mechanism for displacement generation. The displacement sensor beam (show on Figure 7) with center hole mounting the nanoposition LVDT displacement senor measuring the in-plane deflections became the load cell. Figure 1 shows the schematic of the test chip.

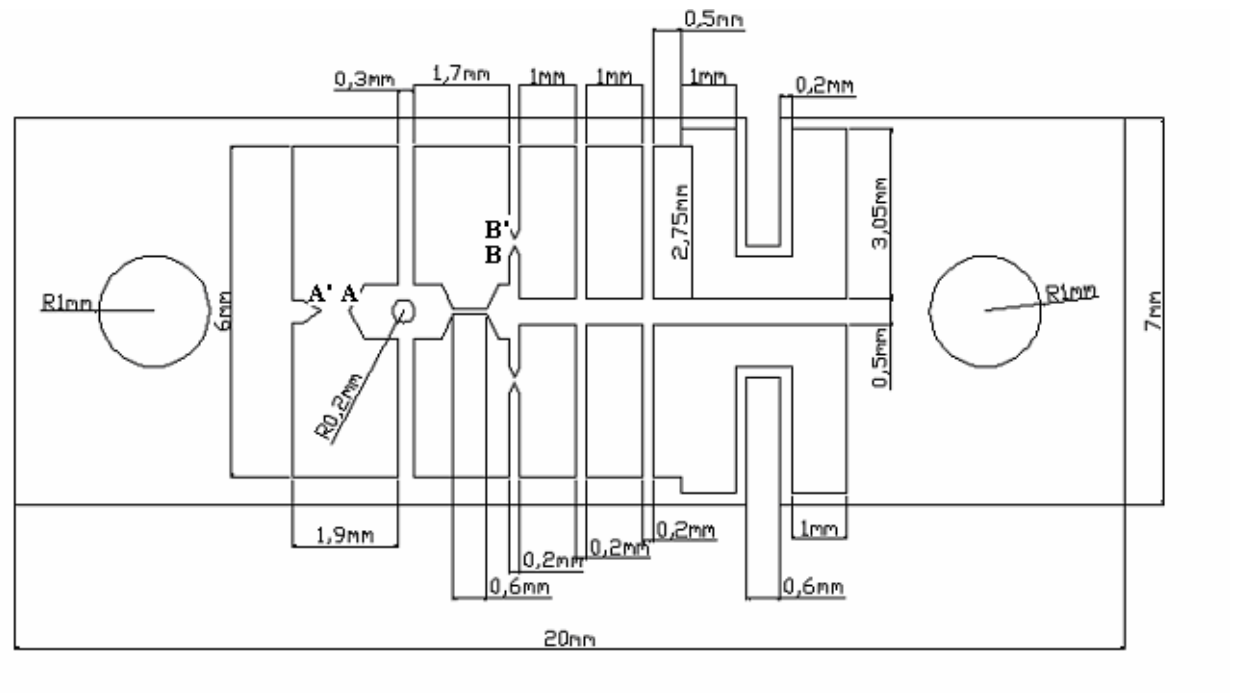

Figure 1: The top view of the test chip.

The central part of the testing chip is the deposited freestanding beam thin film specimen (with a gauge section measuring 600 μm in length and 100 μm in width, and preferred thickness of 200 nm ~ 1 μm) to be tested for its mechanical properties, held at one end by the displacement sensor beam beam with center hole enable to mount a Polytech PI High resulution nanoposition LVDT load cell (Full range of 100μm and a resolution of 10nm) displacement senor underneath, which allows the measurement of in-plane deflections. The other end of the specimen is held by a set of supporting springs, including two supported beams frame and U shape spring. Same thing as other small scale tension tests, the gripping of the specimen is due to the adhesion between the copper frame substrate and the specimen materials. This eliminates the necessity of an extra gripping mechanism for the specimen. The U-shaped springs maintain structural integrity between the fixed and the moving ends of the chip. There are two pin holes on either side of the chip, which are designed to fit two mounting pins. One end fixed and the other can carry the load.

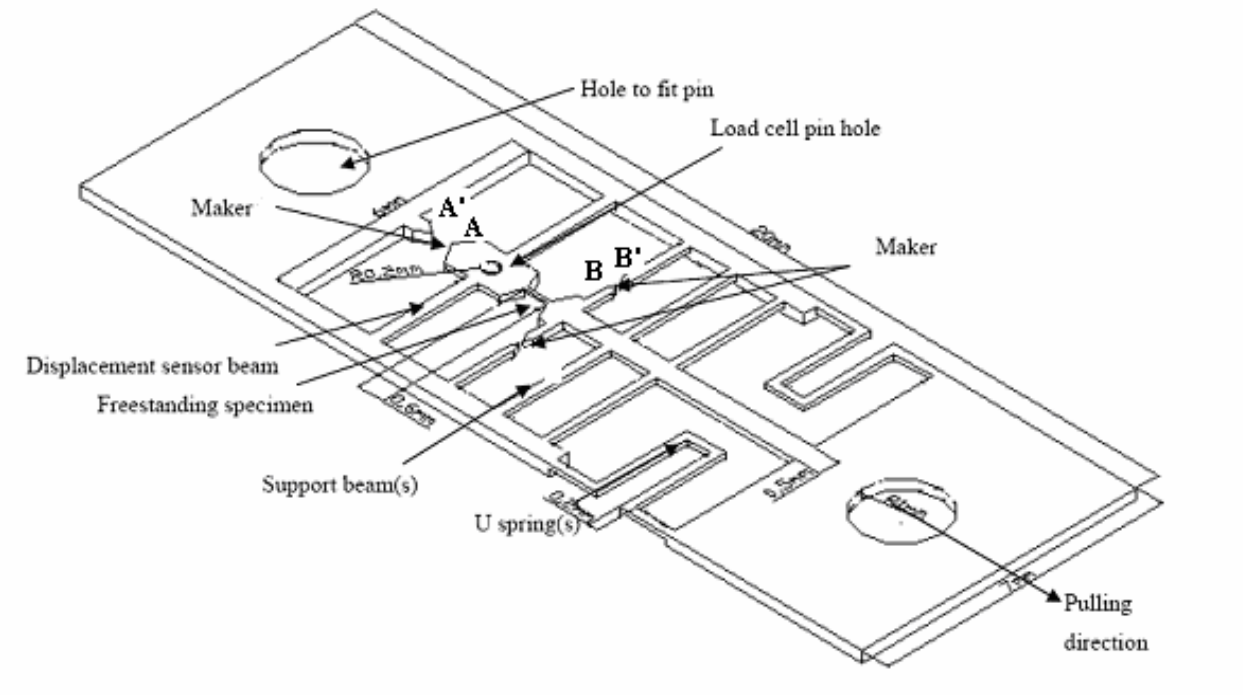

Figure 2: The schematic of the test chip.

The sample was placed flat, with thin film up (copper frame down), on horizontal mounting blocks, or mounts. Stainless steel pins rise vertically out of the top faces of the mounts, fitting through the holes that extend through the copper frame at either end of the tensile sample. The pins fit in the holes. The mounts are held in position relative to one another and the two mounts could move independently of one another.

Tensile force is applied on the specimen by imposing a displacement on one end of the chip, while the other end is held fixed. The displacement is transmitted to the displacement sensor beam by the specimen and causes a deflection on the displacement sensor beam. The force ($F$) on the film specimen is evaluated as $F = k\delta$, where $k$ is the spring constant of the force sensor beam, and $\delta$ is the beam deflection, also shown as the difference of marker gap A A', can be measured from the LVDT displacement senor and the spring constant $k$ of the force sensor beam can be calibrated form Load vs the combination of force transmitted mechanism of LVDT load cell displacement senor and the force sensor beam.

The sample is fabricated on a standard 4 inches diameter double polished 250 μm thick silicon wafer. Each wafer consists of 27 test samples after standard clean room fabrication.

The details of the fabrication procedure are outlined

here:

1. RCA clean Si wafer on a double-side-polished 4-inch (diameter) silicon (Si) wafer.
2. Steam furnace growth 1 μm thin Silicon dioxide layer on double-sides as etching barrier and also scarified layer of the sample.
3. Sputtering deposition of tested thin film materials (later patterned to serve as a freestanding specimen) on one side of wafer with thickness from 0.2~0.8 μm.
4. Using standard photolithography techniques with first mask to pattern the microbeam structure on the front side of the wafer.
5. Using standard photolithography techniques with second mask to pattern the wafer and electroplate copper for 20 μm as a spring structure. Then stripping negative photoresist.
6. Using standard photolithography techniques with third mask to pattern and electroplate copper for 30 μm as frame structure. Then stripping negative photoresist.
7. Using buffered HF to etch away silicon dioxide thin layer and separate the copper framed specimens and the silicon wafer.

Figure 3 shows the process schematic (an example of TaN thin film specimen). Figure 4 shows an example of one mask for the photolithograph procedure.

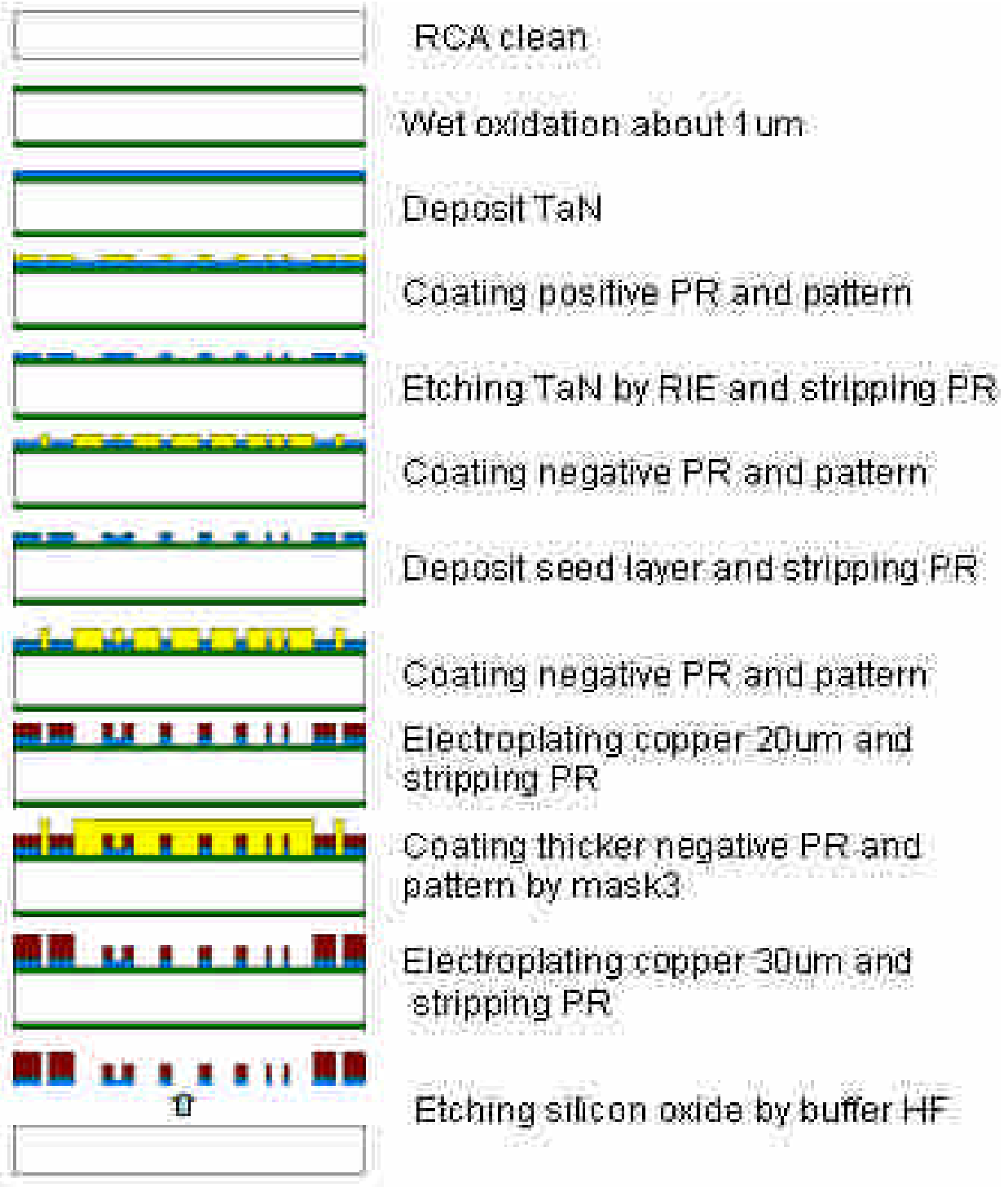

Figure 3: The process schematic of TaN thin film specimens.

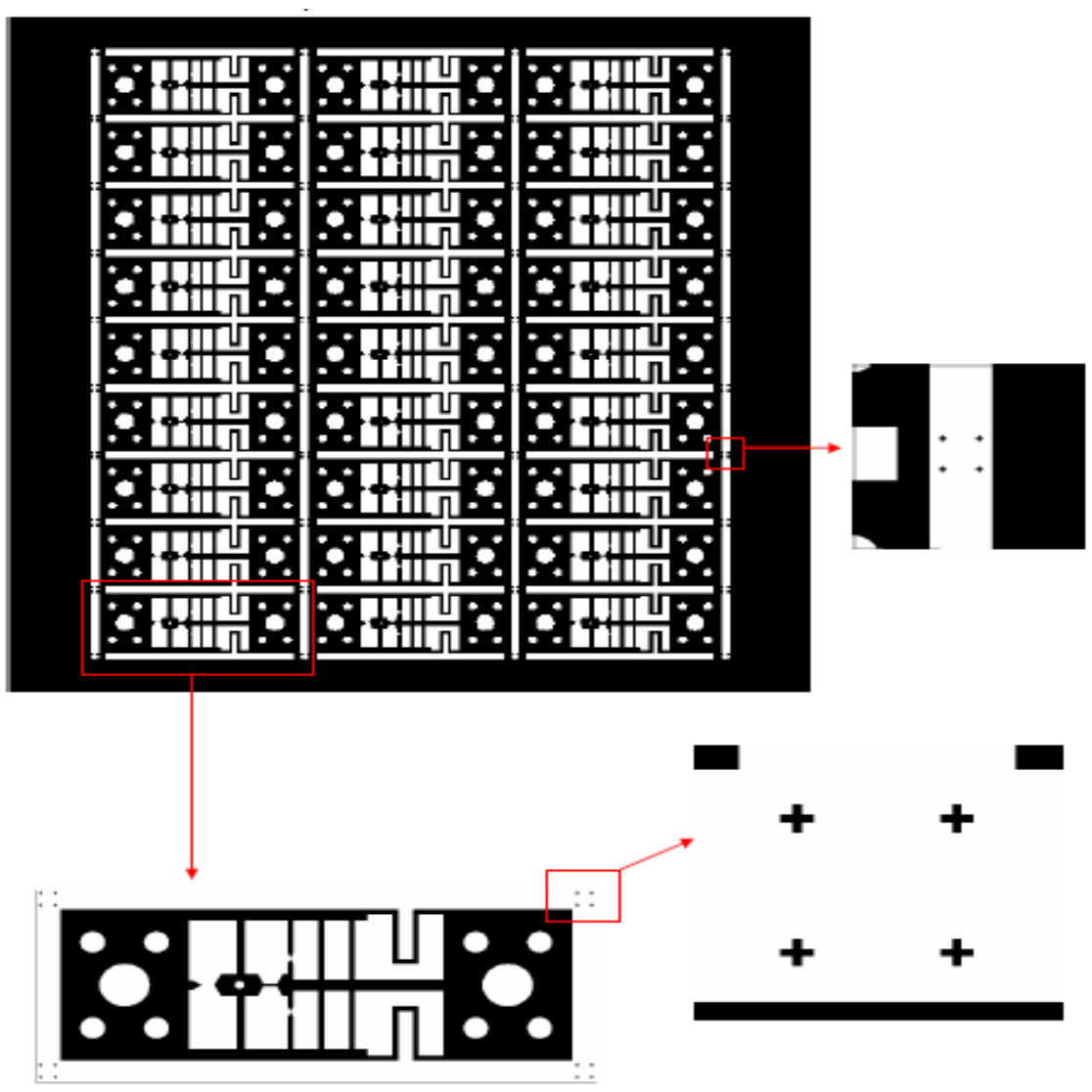

Figure 4: Mask design of the specimens.

The sample fabrication method involves three steps of lithography and two steps of electroplating frame as firm structure to hold freestanding thin films with thickness on the order of 100 nanometers to 1 micron patterned by lift off or chemical etching. Figure 5 shows a fabricated microtensile specimen of 200nm copper thin film ready to be tested.

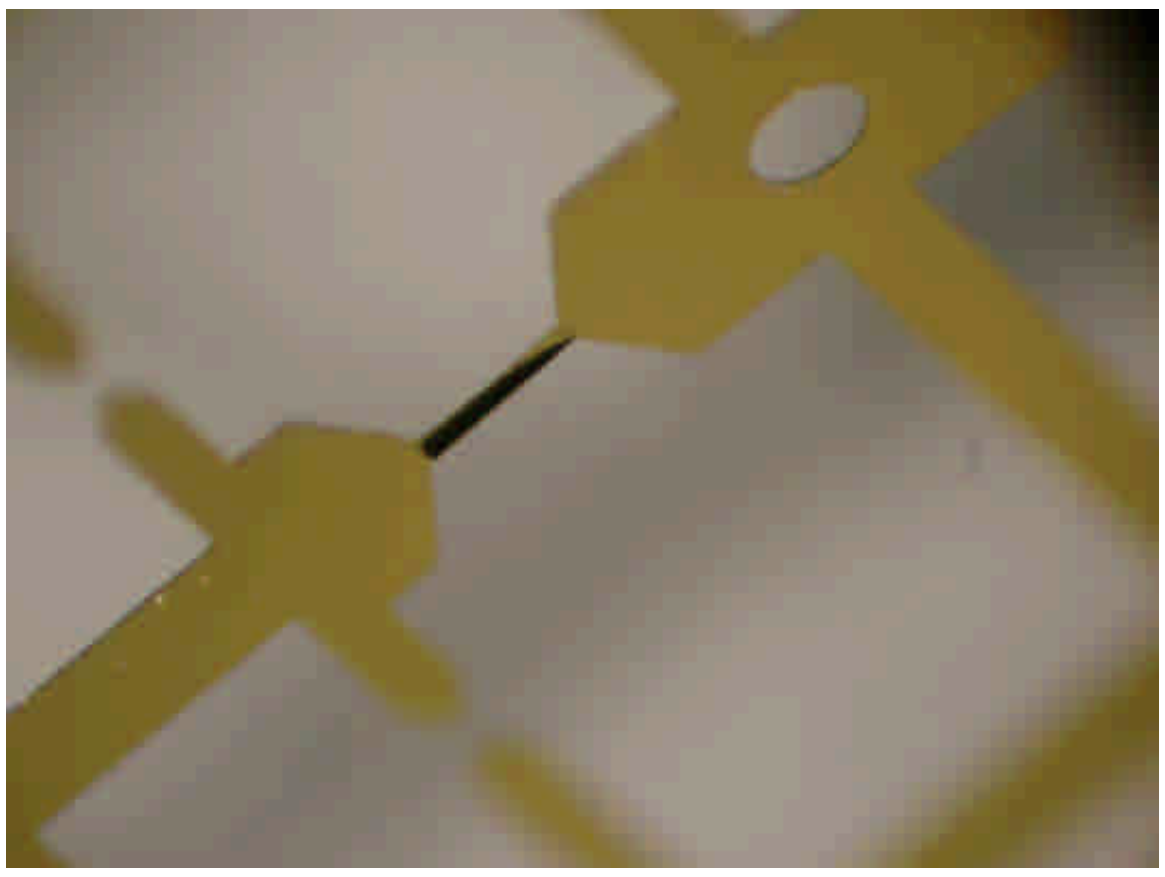

Figure 5: A specimen of 200nm copper thin film.

## 2.2. SYSTEM SET UP

The stage apparatus is custom design and is made from stainless steel equipped with environmental cells and controlled by PC through National Instrument LabVIEW program. In addition, external microscope with CCD camera is equipped to acquire in-situ image for study during testing. The maximum displacement provided by the piezoelectric actuator has travel range over 50 μm so with the gauge length of 600 μm in specimens give us plenty of rooms to measure mechanical behavior over plastic deformation range.

A custom-designed experimental assembly is built to assure well alignment of the sample. This assembly consists of a micromechanical testing system with height-adjustable grip pins, built-in piezoactuator with position sensor, load cell and temperature sensor. The control electronics include a closed-loop piezoelectric controller, amplifier and waveform generator. Monitored signals are conditioned and then fed into an A/D board which is located in a PC. Data acquisition is performed with LabVIEW software. During sample testing, the micromechanical testing system is placed in a thermally insulated box, and the samples are monitored with an optical microscope (Figure 7). A schematic figure of the micromechanical testing system is shown in Figure 6 & 7.

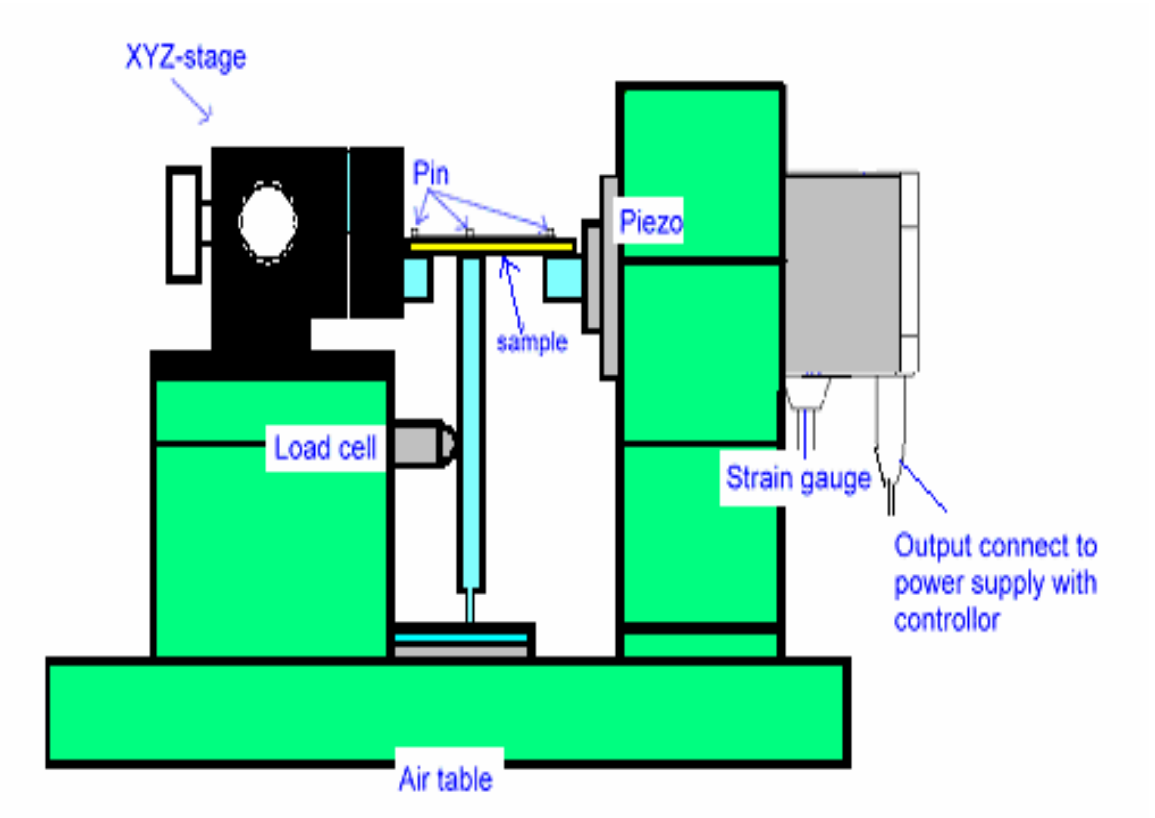

Figure 6: System schematic.

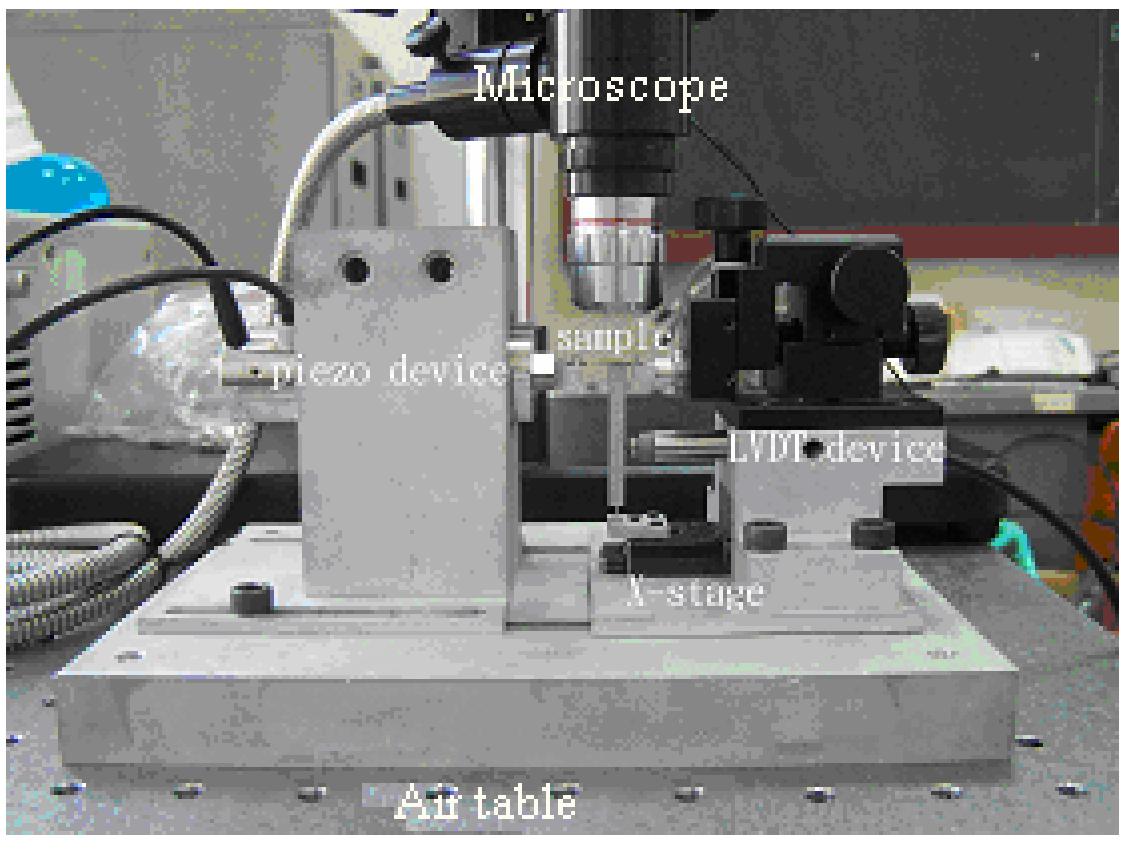

Figure 7: System set up.

Loading was applied using a piezoelectric actuator with 0.1 μm resolution connected through pin hole into the test chip specimen. The designed spring frame (Spring1) inside the test chip was to reduce loading misalign errors by 6 orders of magnitude.

Loads were measured by connected specimens (Spring 2) through a pin connected rod using its bottom pivot to transfer loads and displacement from sample load sensing beam (Spring 3) into a LVDT load cell with a resolution of less than 0.1 mN. The entire testing apparatus was mounted inside a thermally insulated enclosure. This had the effect of limiting thermally induced drift in the system.

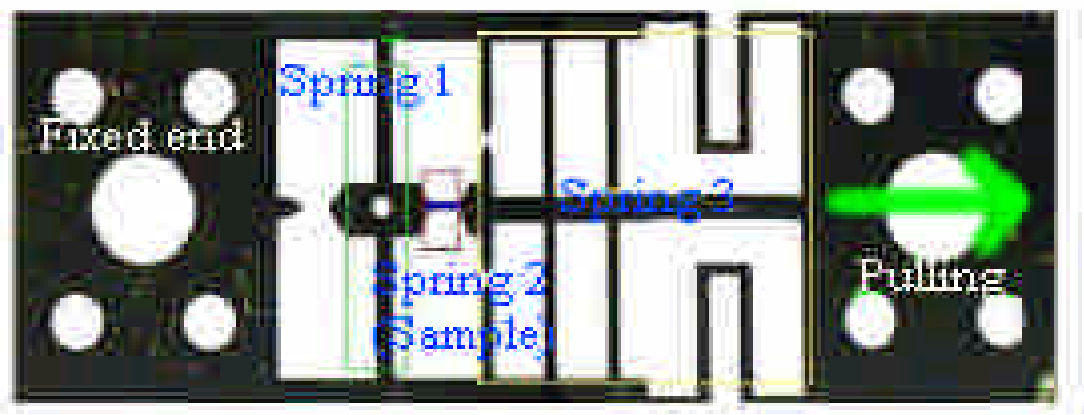

Figure 8: Schematic view of sample testing scenario.

## 3. RESULTS AND DISCUSSION

A typical test history involved constant strain rate straining to a predetermined strain level, a constant strain relaxation period of 20-120 sec duration, then constant strain unloading followed by a zero load hold period. This loading cycle was repeated three to four times. Strain rates for the loading/unloading phases were 3.3 x $10^{-4}$ $sec^{-1}$ or 3.3 x $10^{-3}$ $sec^{-1}$. The displacements vs. time basic test profiles are shown in Figure 9.

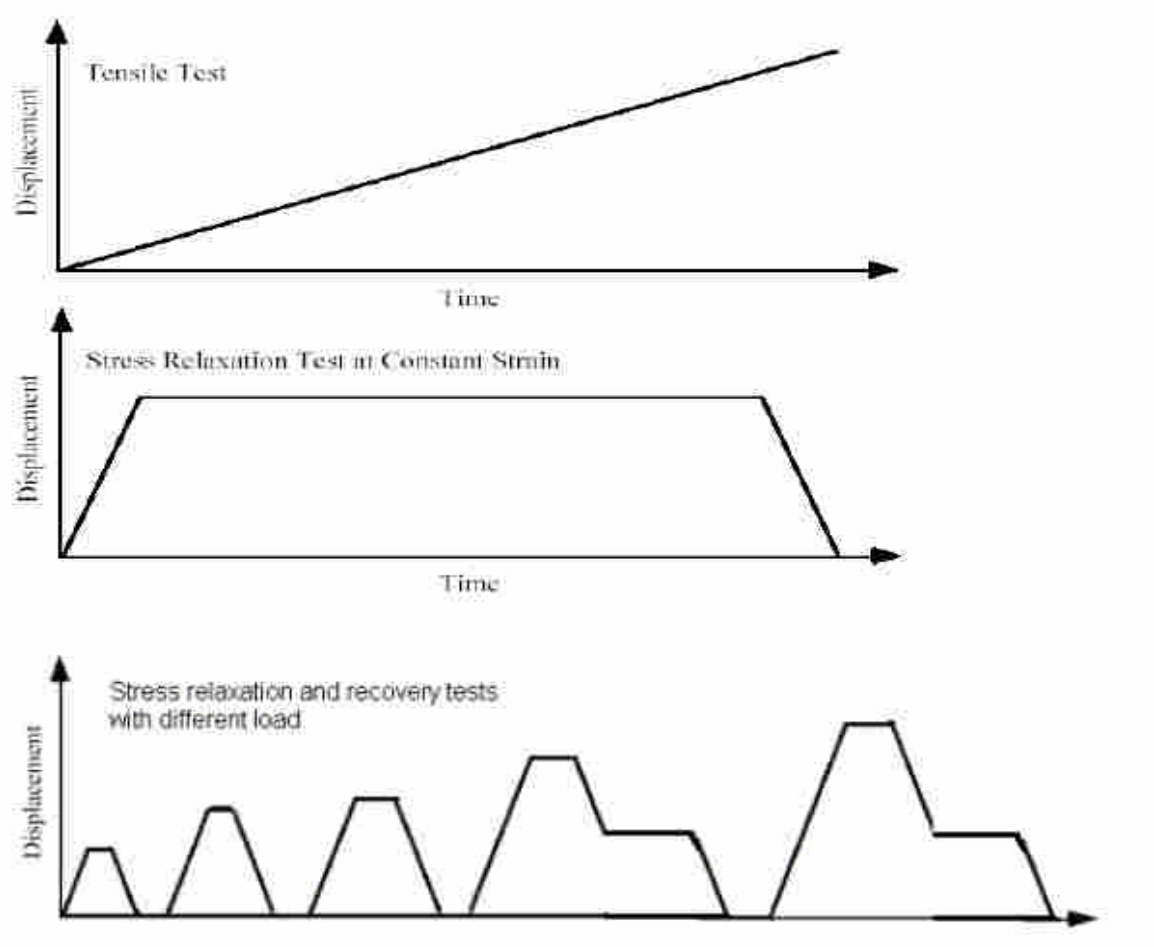

Figure 9: Basic test profiles.

Certain thin films applicable in MEMS were tested including sputtered Chrome, Copper and Tantalum Nitride thin films.

Figure 10 shows one typical stress vs time loading cycle of a tested thin film sample. The stress response is not a straight line but has an upward curvature as strain is increased. The curve becomes more linear at higher stress and strain levels. This is similar to the behavior seen previously for pure Au [10, 11]. We believe the non-linear behavior is largely due to straightening of wrinkles along the edges of the sample and in the body of the sample that is a result of compressive bending which the sample had experienced since it was fabricated. The gradual increase in slope of the stress/strain curve is evident in most of the metal samples we tested.

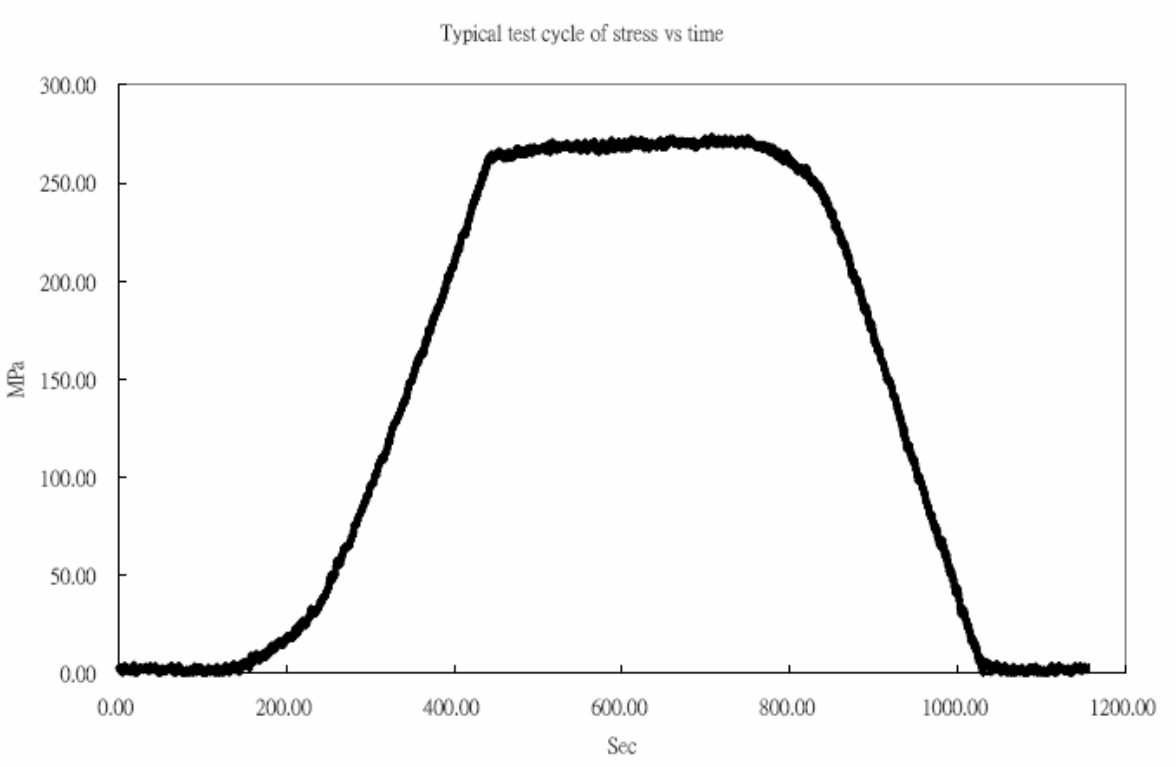

Figure 10: A typical stress vs time loading cycle of test

Figure 11 shows the stress-strain behavior of a 5% Au - Cr film at room temperature. It exhibits simple elastic behavior to yielding point of plastic region at approximately 350 MPa. Linearity was observed in the elastic region which demonstrates the quality of the specimen and the effectiveness of the test instrument.

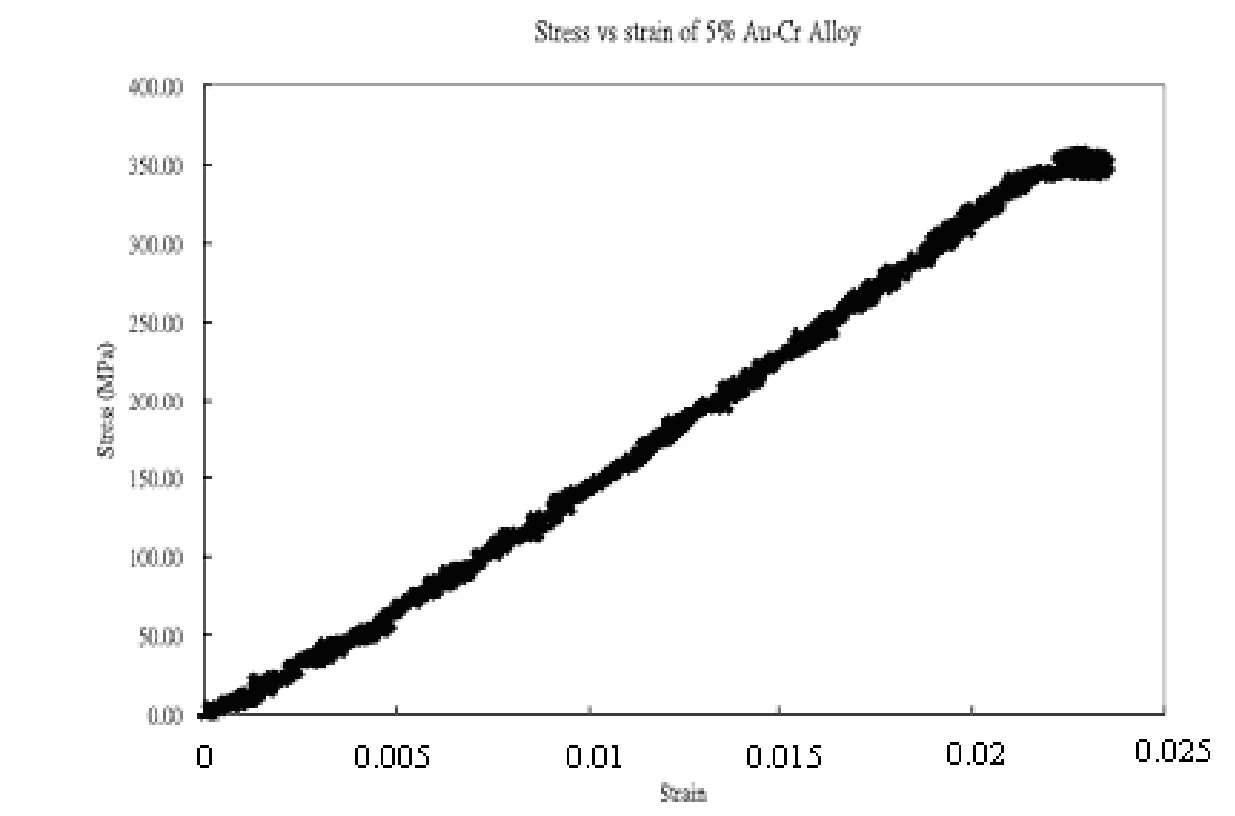

Figure 11 shows the stress-strain behavior of a 5%

The slope of the elastic portion of the curve yielded an average elastic modulus of approximately 161GPa. This value is about 15% less than the modulus of bulk Cr. The difference may reflect a real difference in the elastic properties of the unsupported thin film compared to the bulk but the difference is not much more than the uncertainties in our measurement.

Yield stress is observed by taking 0.2 % straight line to the plastic deformation point of the stress-strain curve shows the value of 350MPa. This is the evidence of the Hall-Petch relation:

$$\sigma_0 = \sigma_i + kD^{-1/2}$$

Where *K* is a constant and *D* is the mean grain size, that the grain size dependence is related to the length of a slip band, and that the maximum slip band length is determined by the grain size.

The modulus values that are listed in Table I were extracted from the data throughout all the tested of each individual samples. The reported uncertainties represent the maximum deviation from the mean observed over several tests of nominally identical beams.

| | **Cu 200nm** | **Au-Cr 200nm** | **TaN 400nm** | **TaN 800nm** |
|---|---|---|---|---|
| **E (GPa)** | **120** | **161** | **151** | **259** |

The values are about 15% less than the bulk moduli. It is not clear the reduction of the modulus in due to size effect. However, the results are closely similar to the previous study of Saif [4,5]、Nix[9]、 Richard[10,11]. We suggest the wrinkles along the edges of the sample and in the body of the sample experienced since it was fabricated is a result of major factor. The concrete understanding of this behavior is not yet complete.

## 4. CONCLUSION

We have demonstrated a custom-designed microtensile apparatus capable of carrying out tests on sub-micro scale thin films for MEMS applications. In this paper, we have shown a few examples of measurements made with it. A more complete account of the results will be further detailed in a future paper.